\documentclass[final,5p,times,twocolumn]{elsarticle}
\usepackage[colorlinks,citecolor=blue,linkcolor=blue,anchorcolor=blue,filecolor=blue,urlcolor=blue]{hyperref}
\usepackage{amssymb,amsmath}
\usepackage{CJKutf8}
\usepackage{lipsum}
\usepackage{setspace} 
\usepackage[authormarkup=none]{changes}
\definechangesauthor[name=RA,color=red]{RA}
\biboptions{sort&compress}
\makeatletter
\def\NAT@def@citea{\def\@citea{\NAT@separator}}
\makeatother

\journal{Physics Letters B}
\begin{document}
\begin{CJK*}{UTF8}{gbsn}
\begin{frontmatter}
\title{Microscopic study of quasifission dynamics in hot fusion reactions for synthesizing superheavy nuclei with $Z=112$--120}
\author[first]{Xiangquan Deng~(邓祥泉)}
\author[first]{Lu Guo~\corref{cor}(郭璐)}
\ead{luguo@ucas.ac.cn}
\cortext[cor]{Corresponding author at: School of Nuclear Science and Technology, University
of Chinese Academy of Sciences, Beijing 101408, China.}
\affiliation[first]{organization={School of Nuclear Science and Technology, University of Chinese Academy of Sciences},
            city={Beijing 101408},
            country={China}}

\begin{abstract}
In the synthesis of superheavy element (SHE) via heavy-ion fusion reactions, 
quasifission is one of the major factors hindering superheavy nuclei (SHN) formation and the mechanism behind this process is intricate.
We investigate dynamics of quasifission in hot fusion reactions synthesizing SHN with $Z=112$--120 using microscopic time-dependent Hartree-Fock theory in a total of 18 reactions.
Remarkably, the nucleon numbers of heavy fragments distribute closely around certain quantum shells in these reactions, 
highlighting the crucial role of shell effects in fragment formation.
In the reactions with $^{48}$Ca, $^{45}$Sc, $^{50}$Ti and $^{51}$V projectiles, the formation of heavy fragment is dominantly driven by the double spherical shells of $^{208}$Pb.
In contrast, the influence of the double octupole deformed shells at $Z=88$ and $N=136$ is more pronounced in $^{54}$Cr-induced reactions, 
resulting in a tendency of producing pear-shaped $^{224}$Ra heavy fragment.
Moreover, in the reaction with a heavier projectile, the colliding system tends to undergo a more rapid quasifission.
This may be responsible for significantly reduced fusion probability and synthesis cross section observed in the reactions with projectiles heavier than $^{48}$Ca.
These results elucidate quasifission mechanisms behind the reactions for synthesizing new SHEs $Z=119$ and $Z=120$. 
\end{abstract}

\end{frontmatter}
Synthesis of new superheavy element (SHE) is currently a frontier area in nuclear physics~\cite{Oganessian2015_RPP78-036301,Giuliani2019_RMP91-011001,Smits2023_NRP6-86}. 
Superheavy nuclei (SHN) can be populated via heavy-ion fusion reactions with different projectile-target combinations.
The SHEs $Z=114$--118 were synthesized by using the hot fusion reactions with $^{48}$Ca projectile and various actinide targets~\cite{Oganessian2004_PRC69-054607,Oganessian2004_PRC70-064609,Oganessian2006_PRC74-044602,Oganessian2007_PRC76-011601R,Oganessian2013_PRC87-014302,Oganessian2013_PRC87-054621}. 
The heaviest SHE at present Og ($Z=118$) was synthesized in 2006~\cite{Oganessian2006_PRC74-044602} and the most recent production of a new SHE occurred in year 2010---the isotopes $^{293,294}$Ts of element $Z=117$ were produced with the combination $^{48}$Ca + $^{249}$Bk~\cite{Oganessian2013_PRC87-054621}. 
However, more than a decade has passed since the discovery of the last new SHE $Z=117$ and the synthesis of SHEs beyond Og is still a great challenge due to extremely low production cross sections.
In the reactions, quasifission is the major process competing with fusion, thereby it stands as one of the primary factors hindering the formation of SHN~\cite{Hinde1995_PRL74-1295,Yu2017_SCPMA60-092011,Guo2018_PRC98-064609,Hinde2018_PRC97-024616,Tanaka2021_PRL127-222501}.
In quasifission, the colliding partners reseparate after interacting for a few to several tens of zeptoseconds with energy dissipation and nucleon transfer~\cite{Hanappe1974_PRL32-738,Simenel2020_PRL124-212504}.
Given its high complexity, an in-depth study of the quasifission process is pivotal for advancing the synthesis of new SHEs.

Characteristics of quasifission have been investigated by plenty of experiments.
Many factors---the static deformation, orientation and shell structure of the colliding nuclei, the incident energy---were confirmed to have strong impact on the reaction outcome~\cite{Hinde2021_PPNP118-103856}.
Aside from the experimental study, various models and approaches, which are phenomenological or microscopic, are employed to describe quasifission process and predict the observables. 
In the phenomenological models, such as the dinuclear system model~\cite{Adamian2003_PRC68-034601,Zhu2014_PRC89-024615,Deng2023_PRC107-014616} and the multidimensional Langevin equations~\cite{Zagrebaev2005_JPG31-825,SiwekWilczynska2019_PRC99-054603}, 
quasifission is treated based on distinctly different assumptions about the colliding process.
Microscopic methods, such as the time-dependent Hartree-Fock (TDHF) theory~\cite{Nakatsukasa2016_PRC88-045004,Simenel2018_PPNP103-19,Stevenson2019_PPNP104-142,Sun2022_CTP74-097302}, 
the quantum molecular dynamics model~\cite{Zhao2015_PRC92-024613,Wang2016_PLB760-236} and the Boltzmann-Uehling-Uhlenbeck model~\cite{Feng2023_PRC107-044606}, 
simulate heavy-ion reactions at nucleon level.
Especially, the TDHF theory provides a fully microscopic and self-consistent framework for revealing reaction mechanisms~\cite{Negele1982_RMP54-913,Simenel2007_PRC76-024609,Guo2007_PRC76-014601,Reinhard2007_EPJA32-19,Umar2008_EPJA37-245,Guo2008_PRC77-041301R}.
It gives reliable descriptions on intricate processes such as fusion~\cite{Guo2012_EWoC38-9003,Simenel2013_PRC88-024617,Shi2017_NPR34-41,Guo2018_PLB782-401,Sekizawa2019_PRC99-051602,Godbey2019_PRC100-054612,Sun2023_PRC107-L011601,Gumbel2023_PRC108-L051602,Sun2023_PRC107-064609}, fission~\cite{Simenel2014_PRC89-031601R,Scamps2018_Nature564-382,Huang2024_PRC110-064318,Huang2024_EPJA60-100,Qiang2025_PLB861-139248}, multinucleon transfer~\cite{Sekizawa2013_PRC88-014614,Wu2019_PRC100-014612,Ayik2020_PRC102-024619,Wu2020_SCPMA63-242021,Jiang2020_PRC101-014604,Wu2022_PLB825-136886}
and quasifission~\cite{Umar2015_PRC92-024621,Li2019_SCPMA62-122011,Godbey2019_PRC100-024610,Li2024_PRC110-064607,Simenel2024_EWoC306-01019}.

\begin{figure*}[t]
  \centering
  \includegraphics[width=1\textwidth]{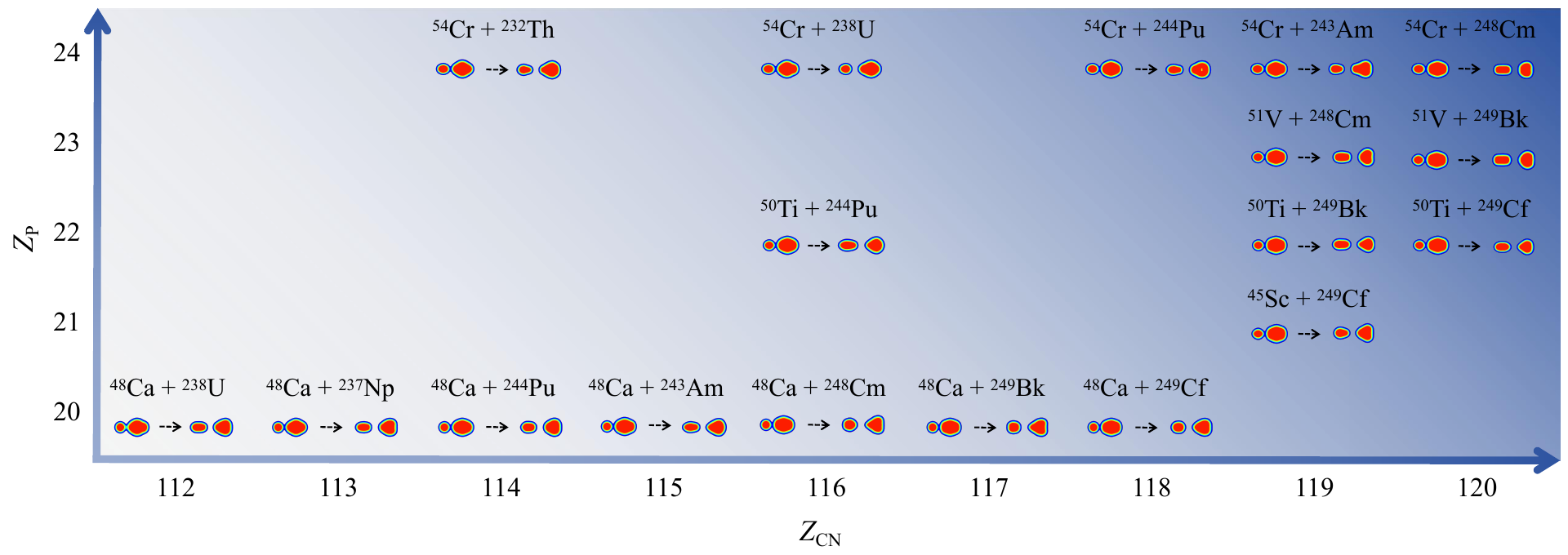}
  \caption{Reactions systematically studied in this work.
  $Z_\mathrm{P}$ and $Z_\mathrm{CN}$ are charge numbers of projectile and compound nucleus, respectively.
  For each reaction, we present the density distribution at the touching configuration and the distribution of quasifission fragments.
  The density profiles are obtained in the TDHF calculations with the effective interaction SLy4d~\cite{Kim1997_JPG23-1267}.
  In the simulation of each reaction, the collision is central and initial orientation of the reactants is tip-tip.
  The incident energy is corresponding to a compound nucleus excitation energy around 30 MeV.}
  \label{fig:1}
\end{figure*}

Among various factors influencing quasifission process, quantum shell effects have significant impact on the dynamic evolution of colliding system. In experiments, the shell effects associated with $^{208}$Pb were manifested by peaks in mass and charge distributions of quasifission fragments~\cite{Morjean2017_PRL119-222502}. 
Meanwhile, multiple TDHF studies reveal spherical shell closures of $^{208}$Pb driving fragment formation in individual reactions such as $^{40}$Ca + $^{238}$U~\cite{Wakhle2014_PRL113-182502},
$^{64}$Ni + $^{238}$U~\cite{Sekizawa2016_PRC93-054616} and $^{48}$Ca, $^{50}$Ti + $^{249}$Bk~\cite{Umar2016_PRC94-024605,Li2022_PLB833-137349}. 
Besides the $^{208}$Pb spherical shells, the role of octupole deformed shell effects in fragment formation was first revealed in fission studies~\cite{Scamps2018_Nature564-382}.
The fission work indicates that octupole deformed shells may also significantly influence the fragment formation in heavy-ion collisions~\cite{Simenel2021_PLB822-136648,Lee2024_PRC110-024606,Wu2026_PRC113-024618}.
The partial mass equilibration in $^{50}$Ca + $^{176}$Yb might be attributed to the octupole deformed shells at $Z=52$ and 56~\cite{Lee2024_PRC110-024606}.
In the inverse quasifission of $^{160}$Gd + $^{176}$W, the octupole deformed shell at $N=88$ dominates the formation of light fragment within an incident energy range~\cite{Wu2026_PRC113-024618}.

However, the theoretical studies above emphasize the impact of shell effects on quasifission in several reactions and current understanding of shell effects in quasifission remains primarily case-specific.
No systematic study has been conducted to convincingly demonstrate that the shell effects universally govern quasifission in hot fusion systems.
Notably, experimental studies of a series of $^{48}$Ti-induced reactions revealed no significant shell-related signatures in quasifission mass spectrum, raising doubts about the role of shell effects in quasifission dynamics~\cite{Hinde2022_PRC106-064614}.
It remains unclear whether shell-driven fragment formation constitutes a universal mechanism or merely reflects the unique dynamics of specific target-projectile combinations.
On the other hand, dramatic decreases in fusion probability and synthesis cross section have been experimentally confirmed, as the projectile changes from $^{48}$Ca to $^{50}$Ti and $^{54}$Cr~\cite{Banerjee2019_PRL122-232503,Gates2024_PRL133-172502,Oganessian2025_PRC112-014603,Oganessian2026_PRC113-014614}.
Such decreases may be related to notably different quasifission behavior in the reactions.
These unresolved issues motivate systematic studies on reactions with $^{48}$Ca and heavier projectiles $^{50}$Ti and $^{54}$Cr, along with in-depth investigations into quasifission mechanisms.

For the first time, 
we systematically investigate the quasifission dynamics of 18 hot fusion reactions with different projectiles using the microscopic TDHF method. 
We study the $^{48}$Ca projectile reactions with which SHN with $Z=112$--118 were experimentally synthesized~\cite{Oganessian2004_PRC69-054607,Oganessian2004_PRC70-064609,Oganessian2006_PRC74-044602,Oganessian2007_PRC76-011601R,Oganessian2013_PRC87-014302,Oganessian2013_PRC87-054621}.
By combining stable isotopes of elements with $Z > 20$ and some long-lived actinide nuclides, 
we study projectile-target combinations applicable for the synthesis of new SHEs $Z=119$ and $Z=120$.
These reaction systems, widely utilized in experiments of multiple nations, demand a comprehensive investigation.
Considering the prevalent use of $^{50}$Ti and $^{54}$Cr in current SHN synthesis experiments, several reactions with these projectiles are studied.
\begin{figure}[t]
  \centering
  \includegraphics[width=0.5\textwidth]{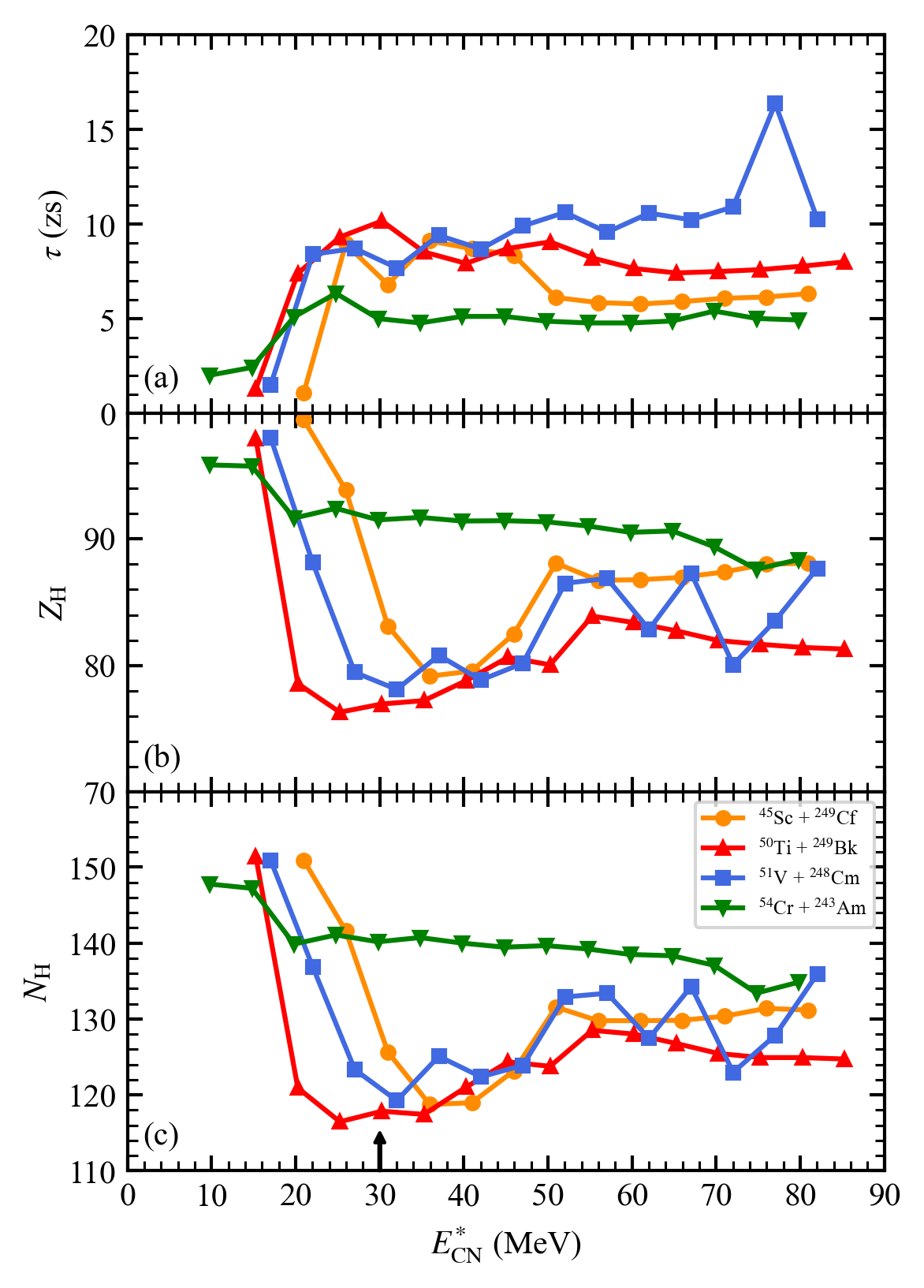}
  \caption{Contact time $\tau$ (a), proton number $Z_\mathrm{H}$ (b) and neutron number $N_\mathrm{H}$ (c) of the heavy quasifission fragment
  as functions of the excitation energy of compound nucleus $E^{*}_\mathrm{CN}$ for the reactions synthesizing superheavy element $Z=119$.
  The energy threshold of 30 MeV is located with a black arrow.}
  \label{fig:2}
\end{figure}
\begin{figure*}[t]
  \centering
  \includegraphics[width=0.95\textwidth]{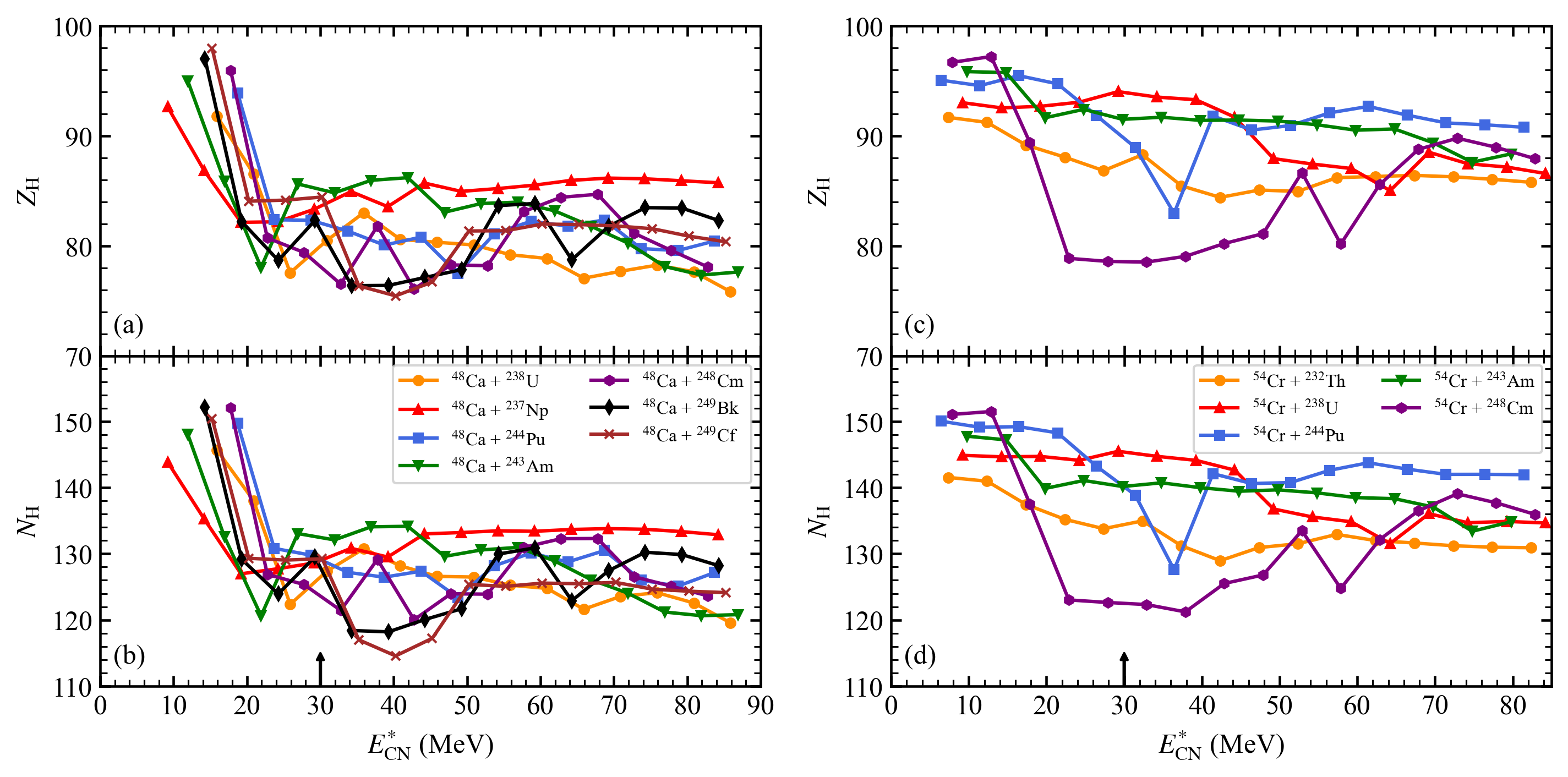}
  \caption{Same as Fig.~\ref{fig:2}(b)--(c), the neutron number distributions for $^{48}$Ca (panels a and b) and $^{54}$Cr (panels c and d) projectile reactions.}
  \label{fig:3}
\end{figure*}

Here we briefly fomulate the theoretical framework of the TDHF method.
The many-body wave function of the system $\Phi(\mathbf{r},t)$ is approximated by a Slater determinant composed of single particle states $\phi_\lambda(\mathbf{r},t)$
\begin{equation}
  \Phi(\mathbf{r},t) = \frac{1}{\sqrt{N!}} \det \bigl[ \phi_\lambda(\mathbf{r},t) \bigr].
\end{equation}
The action of the system is formulated as
\begin{equation}
  S = \int_{t_1}^{t_2} \! dt \, \left\langle \Phi(\mathbf{r},t) | H - i\hbar \partial_t | \Phi(\mathbf{r},t) \right\rangle,
\end{equation}
where $H$ is the many-body Hamiltonian. 
By taking variation of the action with respect to the single-particle states, one obtains the TDHF equations
\begin{equation}
  i\hbar \frac{\partial}{\partial t} \phi_\lambda(\mathbf{r},t) = \hat{h}[\rho] \, \phi_\lambda(\mathbf{r},t).
\end{equation}
This set of nonlinear coupled equations has been solved in three-dimensional coordinate space without any symmetry restrictions.
The TDHF calculations are carried out with an expanded version of Sky3D code as in previous works~\cite{Dai2014_PRC90-044609,Dai2014_SciChinaPMA57-1618,Guo2018_PRC98-064607} where the effective interaction SLy5 was used~\cite{Chabanat1998_NPA635-231,Sun2022_PRC105-034601,Sun2022_PRC105-054610}.
In this study, we adopt the effective interaction SLy4d~\cite{Kim1997_JPG23-1267} for systematic studies of quasifission dynamics.
Some projectile nuclei such as $^{54}$Cr and actinide targets are deformed, necessitating consideration of nuclear orientation effects on the collision dynamics. 
The case where the major axis of the atomic nucleus is aligned along the collision axis is defined as the nucleus being in the tip orientation.
In the calculations, the orientation of the projectile and target is set to be tip-tip and the collisions are central.
Our focus on tip-tip configurations is motivated by their unique dynamical signature. 
It was found that the suppressed nucleon transfer in this orientation (unlike side collisions) makes it a probe for tracking shell-guided dynamics.
To study the energy dependence in collision dynamics, we choose incident energies corresponding to compound nucleus excitation energies $E^{*}_\mathrm{CN}$ below 90 MeV, with an increment of $\Delta E=5$ MeV.
The contact configuration of the colliding nuclei is defined as the state when the minimum density in the neck region exceeds $0.03~\mathrm{fm}^{-3}$.
For each investigated reaction in this work, the density distributions for both the touching configuration and the quasifission fragments are shown in Fig.~\ref{fig:1}.
\begin{figure*}[t]
  \centering
  \includegraphics[width=0.95\textwidth]{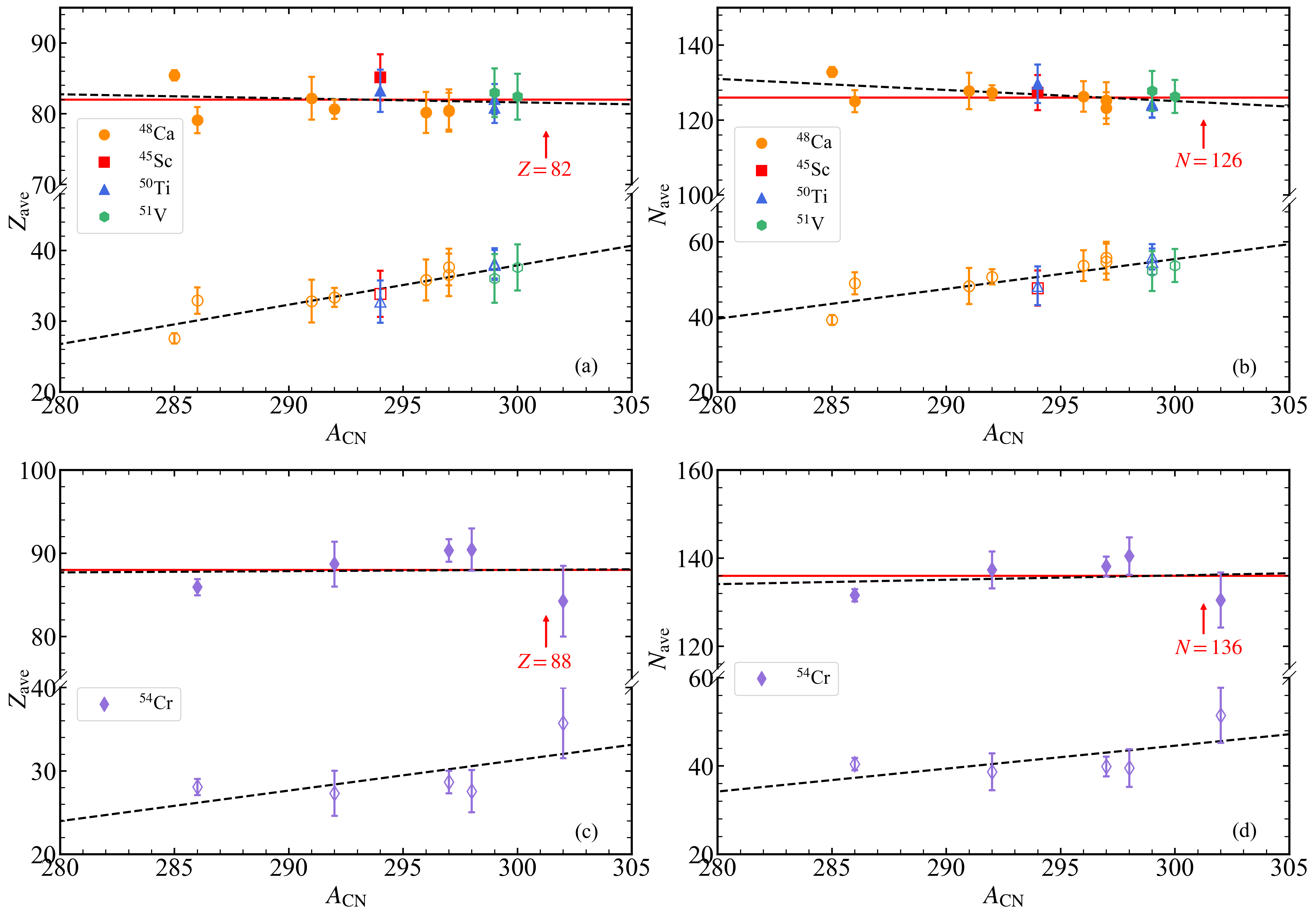}
  \caption{Average proton number $Z_\mathrm{ave}$ and average neutron number $N_\mathrm{ave}$ of light/heavy quasifission fragment for all reactions:
  results for $^{48}$Ca, $^{45}$Sc, $^{50}$Ti and $^{51}$V projectile reactions are shown in (a) and (b);
  results for $^{54}$Cr projectile reactions are shown in (c) and (d).
  For each reaction, the average value and the standard deviation (shown by the error bar) are calculated using the data at $E^{*}_\mathrm{CN} \geq 30$ MeV.
  The abscissa is the mass number of compound nucleus $A_\mathrm{CN}$.
  Results for heavy and light fragment are marked by filled and open symbols, respectively.
  The fitted linear relationships between average nucleon numbers and $A_\mathrm{CN}$ are indicated by black dashed lines. 
  The positions of $Z=82$, $N=126$, $Z=88$ and $N=136$ quantum shells are marked by solid red lines in panels (a)--(d).}
  \label{fig:4}
\end{figure*}
\begin{figure}[t]
  \centering
  \includegraphics[width=0.5\textwidth]{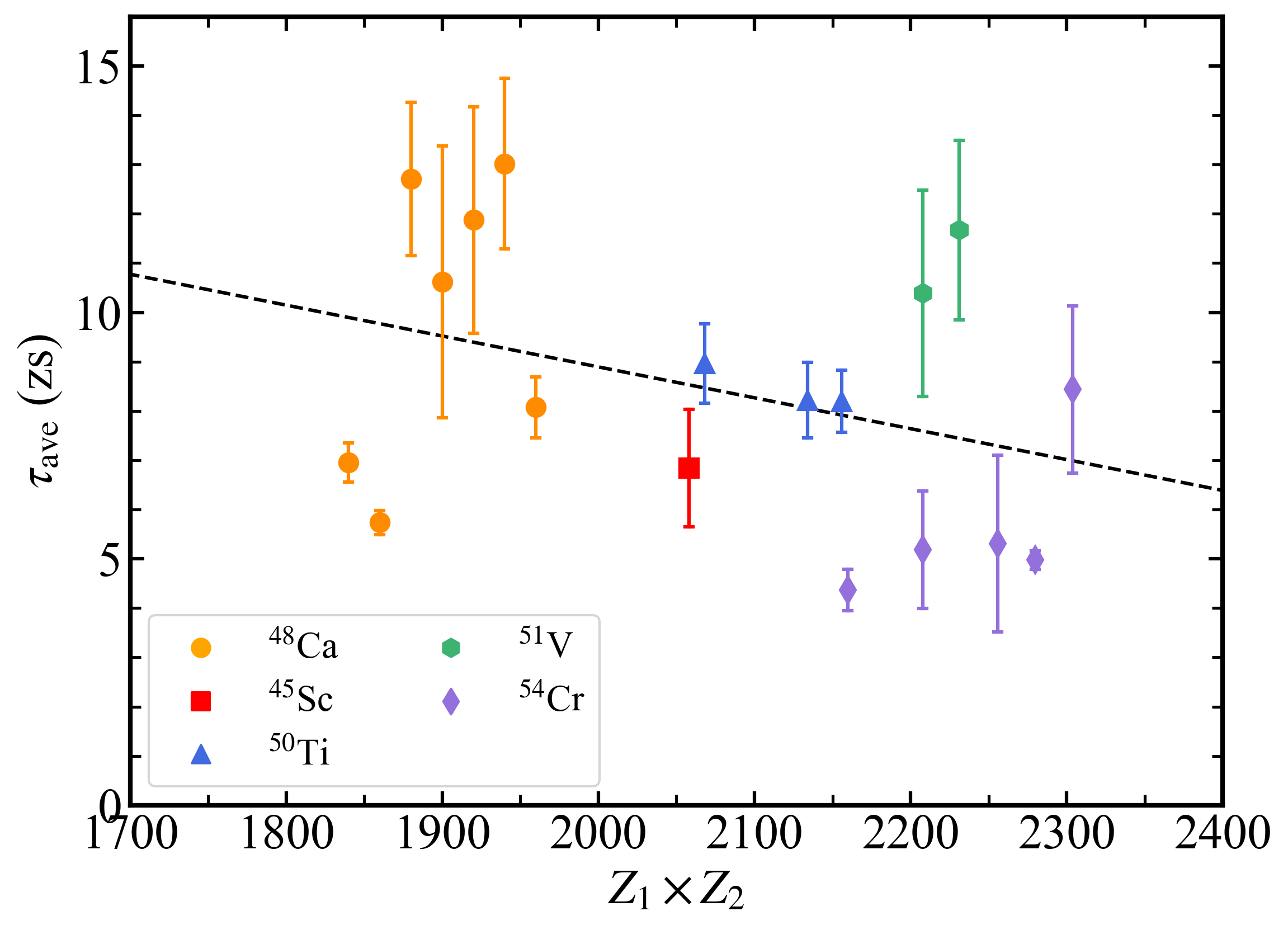}
  \caption{Average contact time $\tau_\mathrm{ave}$ versus Coulomb factor $Z_{1} \times Z_{2}$ of projectile-target combination for all reactions.
  For each reaction, the average value and the standard deviation (shown by the error bar) are calculated using the data at $E^{*}_\mathrm{CN} \geq 30$ MeV.
  $Z_{1}$ and $Z_{2}$ denote charge numbers for projectile and target nuclei, respectively.
  The fitted linear relationship between $\tau_\mathrm{ave}$ and $Z_{1} \times Z_{2}$ is indicated a by black dashed line.}
  \label{fig:5}
\end{figure}

With the TDHF method, we perform large-scale numerical calculations to simulate the quasifission process.
For each reaction, the contact time and the fragment properties are obtained. 
In the following context, we begin by discussing the outcomes of four reactions currently used worldwide to produce SHE $Z=119$. 
To probe the reaction mechanisms, we then compare collision results for systems with the same projectile but different targets, with emphasis on the $^{48}$Ca and $^{54}$Cr projectile reactions.
Finally, for all reactions studied in this work, we investigate the trends in contact time, fragment nucleon numbers and neutron-to-proton ratios, 
thereby revealing and interpreting the quasifission dynamics of these hot fusion reactions.

First, we discuss four reactions that are promising for synthesizing the new SHE $Z=119$: 
$^{45}$Sc + $^{249}$Cf, $^{50}$Ti + $^{249}$Bk, $^{51}$V + $^{248}$Cm, and $^{54}$Cr + $^{243}$Am.  
Figure~\ref{fig:2}(a) shows the contact time $\tau$ as a function of the compound-nucleus excitation energy $E^{*}_\mathrm{CN}$.
The nuclear masses used to convert incident energy to $E^{*}_\mathrm{CN}$ are taken from AME2003~\cite{Audi2003_NPA729-337} for nuclei with experimentally measured values and from FRDM1995~\cite{Moller1995_ADNDT59-185} otherwise.
For hot fusion, the maximal evaporation-residual cross section typically peaks in the $3n$--$5n$ channels, with optimal excitation energies usually exceeding 30~MeV.  
Accordingly, a black arrow marks the 30~MeV threshold in Fig.~\ref{fig:2} and in the following figures.  
All four reactions display a similar energy dependence of the contact time.  
At the incident energies just above the Coulomb barrier (represented by the first data point of each curve), the system undergoes only a shallow contact followed by rapid reseparation, leading to contact times below 3~zs.  
As the energy increases, the contact time grows and eventually approaches a equilibrium value around which it fluctuates.
Such trend reflects the increasingly vigorous quasifission dynamics at higher excitation.  
Above 30~MeV, the $^{54}$Cr + $^{243}$Am reaction exhibits the shortest contact times, around 5~zs, in clear contrast to the other three systems.  
For the reactions $^{45}$Sc + $^{249}$Cf, $^{50}$Ti + $^{249}$Bk and $^{51}$V + $^{248}$Cm, 
we find that more symmetric projectile-target combinations form compound systems with thicker necks after contact. 
The thicker neck delays the rupture of the system, resulting in longer contact times.
Due to the complexity of the collision processes, multiple factors such as nuclear deformation, energy dissipation and relaxation dynamics may interact and influence the evolution of such compound system.

Figures~\ref{fig:2}(b) and~\ref{fig:2}(c) display proton number $Z_\mathrm{H}$ and neutron number $N_\mathrm{H}$ of the heavy quasifission fragment, respectively.
For each system, the nucleon transfer between the reaction partners is limited at the lowest bombarding energy owing to insufficient contact.
Consequently, projectile-like and target-like fragments are produced.
As $E^{*}_\mathrm{CN}$ exceeds 30 MeV, both $Z_\mathrm{H}$ and $N_\mathrm{H}$ exhibit a saturation behavior, 
stabilizing as nearly constant values with further increase in incident energy. 
This trend is consistent with the energy dependence observed in $\tau$, suggesting a common underlying mechanism.
Once the incident energy surpasses the 30 MeV threshold, the dinuclear system achieves a sufficiently long contact time for nucleon exchange. 
At this stage, the net nucleon transfer no longer increases monotonically with energy, 
as the additional excitation energy is primarily redistributed into internal degrees of freedom rather than driving further net mass or charge flow. 
Both the proton and neutron numbers of the heavy fragment fluctuate around their respective saturation values, 
reflecting the statistical nature of the nucleon exchange process at elevated excitation energies.

To further elucidate the reaction mechanisms, we perform a comparative investigation on reactions utilizing identical projectiles but different targets. 
Figures~\ref{fig:3}(a) and~\ref{fig:3}(b) display the nucleon numbers of the heavy quasifission fragments for the $^{48}$Ca projectile systems, and Fig.~\ref{fig:3}(c) and~\ref{fig:3}(d) for the $^{54}$Cr systems.
The comparison of Fig.~\ref{fig:3}(a) and~\ref{fig:3}(c) reveals that across the $^{48}$Ca-induced reactions, the proton numbers of the heavy fragments are distributed within a relatively lower range for energies above the 30~MeV threshold. 
In contrast, the $^{54}$Cr-induced reactions systematically produce heavy fragments with larger proton numbers, as evidenced by the overall distribution of curves.
A similar contrast is also evident in the neutron number distributions of the heavy fragments between Fig.~\ref{fig:3}(b) and~\ref{fig:3}(d).
Such contrasts in nucleon number distributions suggest that the quasifission fragment formation in $^{48}$Ca-induced and $^{54}$Cr-induced reactions are governed by different shell structures. 
The heavy fragments in the $^{48}$Ca-induced reactions are likely influenced by the doubly magic spherical shell closure of $^{208}$Pb, 
which resides near the middle of the observed proton and neutron number ranges. 
On the other hand, the formation of heavy fragments in the $^{54}$Cr systems are plausibly associated with heavier shell structures, 
as reflected in larger fragment proton and neutron numbers. 

In all reactions studied, once the incident energy suffices for deep touch between the reactants, 
both the contact time and heavy-fragment nucleon numbers vary near saturation values with increasing energy. 
However, the results across different systems and energies inevitably contain fluctuations which may obscure the underlying shell effects. 
To reveal which shell structures govern the quasifission process, 
we compute average values of contact time and heavy-fragment nucleon numbers using data at $E^{*}_\mathrm{CN} \geq 30$ MeV. 
In the following context, we discuss systematic results of fragment nucleon numbers, contact time, and neutron-to-proton ratio differences of the 18 reactions.

We observe that the heavy-fragment nucleon numbers in reactions with $^{45}$Sc, $^{50}$Ti, 
and $^{51}$V projectiles lie within nearly the same range as those in the $^{48}$Ca-induced reactions, 
suggesting that the quasifission process in these systems might be governed by the same quantum shells.
Figures~\ref{fig:4}(a) and~\ref{fig:4}(b) show the average proton number $Z_\mathrm{ave}$ and average neutron number $N_\mathrm{ave}$ of the heavy and light fragments in these reactions, 
plotted against the compound nucleus mass number $A_\mathrm{CN}$. 
Heavy-fragment results are marked by filled symbols and light-fragment results by open symbols. 
For each reaction, the standard deviation of the averaged data is shown by the error bar.
If the formation of heavy quasifission fragments in these systems is influenced by the same shell effects, 
linear fitting of the average nucleon numbers across different systems can help identify these shells. 
The resulting linear fits are shown as black dashed lines, 
while the positions of the $Z=82$ and $N=126$ shell closures are marked by solid red lines for reference.
Remarkably, the systematics hold very well across all these reactions. 
The fitted lines for the heavy fragment's proton and neutron numbers align closely with the red reference lines, 
indicating that the proton numbers distribute around $Z=82$ and the neutron numbers around $N=126$. 
This behavior arises from the spherical doubly magic nucleus $^{208}$Pb. 
In contrast, the nucleon numbers of the light fragments increase linearly with $A_\mathrm{CN}$, 
as they consist of the remaining nucleons after the formation of the $^{208}$Pb-like heavy fragment.
We conclude that, in tip-oriented collisions, hot fusion systems induced by $^{48}$Ca, $^{45}$Sc, $^{50}$Ti, and $^{51}$V projectiles share a common quasifission mechanism. 
This universality is driven by the strong tendency to form $^{208}$Pb as the heavy fragment, 
governed by the double shell closures at $Z=82$ and $N=126$.

To identify the shells that dominate the quasifission process of $^{54}$Cr-induced reactions, we calculate the average proton and neutron numbers of the heavy fragments, as presented in Fig.~\ref{fig:4}(c) and~\ref{fig:4}(d). 
We again employ linear fitting to reveal the active quantum shells, with the resulting fits indicated by black dashed lines. 
The positions of $Z=88$ and $N=136$ are marked with solid red lines, which coincide with the experimentally confirmed octupole deformed nucleus $^{224}$Ra~\cite{Gaffney2013_Nature497-199}.  
The fitted lines closely match these red references, demonstrating that the quasifission process in $^{54}$Cr-induced reactions is strongly governed by the double octupole deformed shells at $Z=88$ and $N=136$.
It is noteworthy that Li et al.~\cite{Li2026_PRC114-014626}, 
while employing the same theoretical framework as this work but with the SLy5 effective interaction, 
also concluded that quasifission in tip-tip collisions of $^{54}$Cr + $^{243}$Am is dominated by these octupole deformed shells within a specific energy window. 
Although the present study adopts the SLy4d effective interaction, both studies yield identical shell effects, indicating that the manifestation of shell effects might not depend on the employed effective interaction.

It was revealed that fragment production in both fission and quasifission processes is influenced by similar shell effects~\cite{Simenel2021_PLB822-136648}. 
The quasifission dynamics of the reactions can be analyzed through the compound nucleus potential energy surface (PES)~\cite{Lee2024_PRC110-024606,Simenel2025_NPA1062-123170}.
In Ref.~\cite{Lee2024_PRC110-024606}, the TDHF trajectories of $^{50}$Ca + $^{176}$Yb and $^{96}$Zr + $^{130}$Sn systems were mapped onto the $Q_{20}-Q_{30}$ PES of $^{226}$Th and the quasifission trajectories of both systems are close to the asymmetric fission valley of $^{226}$Th. 
Based on the landscape of the compound nucleus PES, the evolution of the reaction system during the quasifission process can be understood.
Although not directly employed in this work, the approach provides inspiration, 
enabling qualitative discussion of collision outcome differences across projectile-target combinations.
We expect that the PES of the $^{54}$Cr-induced compound nucleus develops a valley that leads the system towards the pear-shaped $^{224}$Ra nucleus.
In contrast, for $^{48}$Ca-induced reactions, the PES may exhibit a deep valley directed towards the spherical doubly magic $^{208}$Pb,
which accounts for the trends shown in Figs.~\ref{fig:4}(a) and~\ref{fig:4}(b). 
In the $^{48}$Ca reactions, the formation of the heavy fragment is driven by the strong spherical closures, 
while in the $^{54}$Cr reactions it is driven by the octupole deformed shells, causing the quasifission trajectories to deviate towards the $^{224}$Ra region.
These speculations can further be verified by computing the potential energy surfaces of the compound nuclei.
Moreover, shell effects at $Z=52$--56 influence nucleon distributions in both fission and quasifission~\cite{Simenel2021_PLB822-136648}. 
By analogy, it is plausible that the $Z=88$ and $N=136$ shells could also affect the fission of some superheavy nuclei, leading to asymmetric mass splits.

$^{48}$Ca-induced hot fusion has been successfully used to synthesize superheavy elements, 
but target-material shortages limit its extension to $Z=119$ and $120$, 
motivating the use of heavier projectiles such as $^{50}$Ti and $^{54}$Cr.  
In reactions populating $Z=116$ isotopes, the $4n$ channel cross section for $^{48}$Ca + $^{248}$Cm reaches $3.3^{+2.5}_{-1.4}$ pb~\cite{Oganessian2004_PRC70-064609}, 
while for $^{50}$Ti + $^{244}$Pu it is $0.44^{+0.58}_{-0.28}$ pb~\cite{Gates2024_PRL133-172502}.  
Recently, at a $^{292}$Lv excitation energy of approximately 40 MeV, cross sections of $0.22^{+0.27}_{-0.15}$ pb for $^{50}$Ti + $^{242}$Pu and $0.036^{+0.046}_{-0.024}$ pb for $^{54}$Cr + $^{238}$U have been measured~\cite{Oganessian2025_PRC112-014603}.  
Thus, relative to $^{48}$Ca projectile, the cross section drops by about one order of magnitude for $^{50}$Ti and by nearly two orders of magnitude for $^{54}$Cr.  
A similar trend is observed in cold fusion, where fusion probabilities with $^{50}$Ti and $^{54}$Cr are much lower than with $^{48}$Ca~\cite{Banerjee2019_PRL122-232503}.
Figure~\ref{fig:5} shows the average contact time $\tau_\mathrm{ave}$ as a function of the entrance-channel Coulomb factor $Z_{1} \times Z_{2}$, 
where $Z_{1}$ and $Z_{2}$ are the charge numbers of the projectile and target, respectively.
$\tau_\mathrm{ave}$ decreases systematically with increasing $Z_{1} \times Z_{2}$, and the black dashed line represents a linear fit to the data.
A larger Coulomb factor implies a stronger repulsive force, which drives the system to reseparate more rapidly and thus shortens the contact time.
As the projectile changes from $^{48}$Ca to $^{50}$Ti and $^{54}$Cr, the increasing $Z_{1} \times Z_{2}$ accelerates the quasifission dynamics, 
which may strongly suppress the formation of a fully equilibrated compound nucleus and cause a dramatic reduction of the fusion probability $P_{\mathrm{CN}}$.
Hence, the pronounced differences in synthesis cross sections observed across these reactions can be understood as a consequence of increasing Coulomb repulsion between the colliding partners.
\begin{figure}[t]
  \centering
  \includegraphics[width=0.5\textwidth]{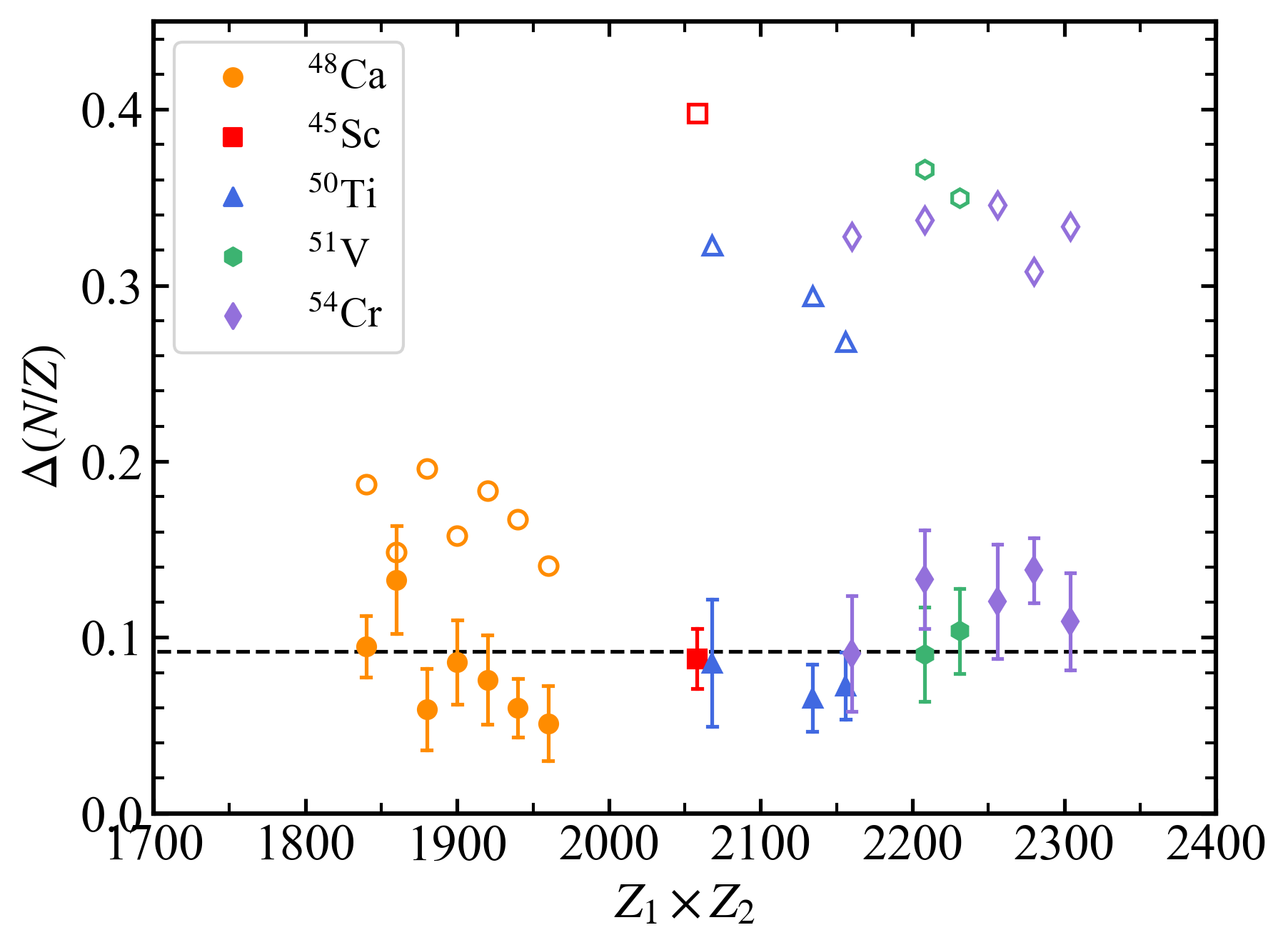}
  \caption{Average value of the neutron-to-proton ratio ($N/Z$) difference between two quasifission fragments $\Delta(N/Z)$, marked by filled symbols. 
  For each reaction, the average value and the standard deviation (shown by the error bar) are calculated using the data at $E^{*}_\mathrm{CN} \geq 30$ MeV.
  The abscissa is the entrance-channel Coulomb factor $Z_{1} \times Z_{2}$.
  The average $N/Z$ difference of all investigated reactions is located by the black dashed line.
  The $N/Z$ differences between the projectile and target nuclei are indicated by open symbols.}
  \label{fig:6}
\end{figure} 

In Fig.~\ref{fig:6}, the average value of the neutron-to-proton ratio ($N/Z$) difference between the heavy and light quasifission fragments $\Delta(N/Z)$ is marked by a filled symbol for each reaction, 
and the $N/Z$ difference between projectile and target is shown by an open symbol.
The entrance-channel Coulomb factor $Z_{1} \times Z_{2}$ is used as abscissa. 
Despite significant differences in avarage contact time, we find neutron-to-proton equilibration is approached during quasifission in all the reactions. 
The average $N/Z$ difference of all 18 reactions is 0.092, 
which is located by a black dashed line in Fig.~\ref{fig:6}.
The $N/Z$ equilibration process, which is more rapid in comparison with mass equilibration~\cite{Simenel2020_PRL124-212504}, may strongly impact the dynamic pathways of quasifission/fusion and diminish the influence of entrance channel on reaction outcomes,
the systematic $\Delta(N/Z)$ values obtained in this work may serve as microscopic constraints for phenomenological models to decribe fusion reactions. 

In summary, this microscopic TDHF research investigates dynamics of quasifission in hot fusion reactions using 
$^{48}$Ca, $^{45}$Sc, $^{50}$Ti, $^{51}$V and $^{54}$Cr projectiles for systhesizing superheavy nuclei with $Z=112$--120. 
Central collisions in tip-tip orientation are simulated across a wide above-barrier energy range.
We find that both contact time of reactants and nucleon numbers of heavy fragments saturate with increasing incident energy in these reactions.
The energy threshold for such equilibrium corresponds to the compound nucleus excitation energy of approximately 30 MeV.

Unexpectedly, the systematic behavior in fragment nucleon number persists remarkably well.
We find that the reactions with $^{48}$Ca, $^{45}$Sc, $^{50}$Ti and $^{51}$V projectiles tend to yield $^{208}$Pb as the heavy fragment.
The quasifission process in these hot fusion systems are governed by an identical mechanism, 
namely the formation of the heavy fragment is dominantly driven by the double spherical shells of $^{208}$Pb.
Surprisingly, the fragment nucleon number distributions in $^{54}$Cr projectile reactions differ significantly from other systems. 
In $^{54}$Cr-induced reactions, the influence of the double octupole deformed shells at $Z=88$ and $N=136$ is more pronounced, 
leading to a tendency in these reactions of producing pear-shaped $^{224}$Ra heavy fragments.
Both the SLy4d and SLy5 effective interactions exhibit identical octupole deformed shell effects that dominate the quasifission process of $^{54}$Cr + $^{243}$Am.

With regard to the reaction dynamics, a heavier projectile systematically shortens the contact time.
As the projectile changes from $^{48}$Ca to $^{50}$Ti and $^{54}$Cr, the increasing entrance-channel Coulomb repulsion accelerates quasifission,
which strongly suppresses the formation of a fully equilibrated compound nucleus. 
This may account for the dramatic reduction in the synthesis cross sections observed experimentally.
Furthermore, the average neutron-to-proton ratio difference between the heavy and light fragments is obtained for each reaction.
These values may serve as microscopic constraints for phenomenological models to decribe fusion reactions and predict quasifission/fusion pathways.
This work reveals the dominant mechanisms of quasifission in hot fusion systems, including the reactions currently used for synthesizing SHEs $Z=119$ and $Z=120$.
The results could potentially benefit future attempts to produce new SHEs.

We should notice that some issues still require further investigations.
This work systematically study the tip-oriented collisions of the hot fusion reactions and discuss the importance of underlying shell effects.
However, previous studies indicate that the shell effects are orientation-dependent.
When deformed projectiles like $^{54}$Cr or targets are involved, 
the reaction process may be affected by orientation of the reactants.
It is of importance to comprehensively investigate the collisions with different projectile-target orientations and impact parameters,
thereby deeply exploring the influence of shell effects.
On the other hand, future studies could further investigate the impact of different effective interactions on quasifission dynamics.

\section*{Acknowledgements}
\begin{spacing}{1.0} 
Xiangquan Deng acknowledges Liang Li for helpful suggestions.
This work has been supported by the Strategic Priority Research Program of the Chinese Academy of Sciences (Grant No. XDB1550100) and
the National Natural Science Foundation of China (Grants No. 12435008, No. 12375127, and No. 12205308).
\end{spacing}
{\tiny
\bibliographystyle{elsarticle-num}
\bibliography{main.bib}
}

\end{CJK*}
\end{document}